\documentstyle[aaspp4]{article}
\begin{document}

\title{Rotation Periods of Late-type Stars in the Young Open Cluster IC 2602}
\author{Sydney A. Barnes\footnote{Currently Lowell Fellow at Lowell Observatory, 1400 W. Mars Hill Rd., Flagstaff, AZ 86001, USA.}$^,$ \footnote{Visiting 
Astronomer at the  Cerro Tololo Inter-American Observatory, operated by the 
Association of Universities for Research in Astronomy (AURA) Inc., under 
contract with the National Science Foundation.}  and Sabatino Sofia }
\affil{Yale University, Astronomy Department, Box 208101, New Haven, CT 06520-8101, USA \\
 email: barnes@lowell.edu, sofia@astro.yale.edu}
\author{Charles F. Prosser$^{2,}$ \footnote{Charles F. Prosser is now deceased.}}
\affil{NOAO, Kitt Peak National Observatory, P.O. Box 26732, Tucson, AZ 85726-6732, USA}
\author{John R. Stauffer }
\affil{Harvard-Smithsonian Center for Astrophysics, 60 Garden Street, Cambridge, MA 02138, USA \\
 email: stauffer@amber.harvard.edu}

\begin{abstract}

We present the results of a monitoring campaign aimed at deriving rotation
periods for a representative sample of stars in the young (30 Myr)
open cluster IC 2602. Rotation periods were derived for 29 of 33 stars 
monitored. The periods derived range from 0.2d (one of the shortest known
rotation periods of any single open cluster star) to about 10d (which
is almost twice as long as the longest period previously known for a cluster
of this age). We are able to confirm 8 previously known periods and derive
21 new ones, delineating the long period end of the distribution. 
Despite our sensitivity to longer periods, we do not detect any variables 
with periods longer than about 10d. 
The combination of these data with those for IC 2391, an almost identical 
cluster, leads to the following conclusions:
\ 1) The fast rotators in a 30 Myr cluster are distributed across the entire
   0.5 $< B-V <$ 1.6 color range.
\ 2) 6 stars in our sample are slow rotators, with periods longer than 6d.
\ 3) The amplitude of variability depends on both the color and the period.
   The dependence on the latter might be important in understanding the
   selection effects in the currently available rotation period database
   and in planning future observations.
\ 4) The interpretation of these data in terms of theoretical models of 
   rotating stars suggests both that disk-interaction is the norm rather
   than the exception in young stars and that disk-locking times range 
   from zero to a few Myr.

\end{abstract}


 \section{Introduction}

IC 2602 is a young (30 Myr-old) open cluster in the Southern Hemisphere 
($\alpha$~=~$10^h41^m$, $\delta$~=~$-64^{\circ}$,~1950). 
At a distance of about 150 pc, it is one of the nearest open clusters.
For the brightest members, Whiteoak (1961) and Braes (1962) provided 
photographic and photoelectric photometry, and spectral types were also 
derived for some of the brighter candidates using low-dispersion spectroscopy 
by Whiteoak (1961) and Abt and Morgan (1972). Braes (1962) also provided 
proper motion membership for some of the bright members of the cluster. 
Despite its proximity, no information was available for the low-mass stars
in the cluster until very recently.

Unfortunately, no proper motion membership is currently available for the
late-type stars in IC 2602, and studies of this cluster have had to overcome
this obstacle in various ways. A ROSAT X-ray survey of IC 2602 reported by
Randich et al. (1995) identified several candidate low-mass members. This work
was followed by additional photometry (Prosser et al. 1996), $v \sin i$
measures and spectroscopic confirmation of memberhip (Stauffer et al. 1997), 
and lithium abundances (Randich et al. 1997)
for these low-mass members and candidate members. 
Rotation period measurements, which are preferred to those of $v \sin i$ 
because they provide the unambiguous measure of a given star's angular 
velocity,  have been provided for a few members of this cluster by
Patten et al. (1996) and  Prosser (1997) but no substantial sample of rotation 
periods has yet been published for this cluster. 

IC 2602 is particularly interesting from the point of view of studying stellar
rotation because it affords a sample of stars that have just descended the
Hayashi track, but have not yet experienced main sequence
angular momentum loss. Thus, they can illustrate the distribution of 
rotational velocities among pre-main sequence stars.
Although data exist for two slightly older clusters, Alpha Per (Stauffer et 
al. 1985, Stauffer et al. 1987, Prosser et al. 1993a, 1993b, 1995, O'Dell and 
Cameron 1993, Bouvier 1996, Prosser and Grankin 1997)
and the Pleiades 
(Van Leeuwen and Alphenaar 1982, Van Leeuwen et al. 1987, Stauffer et al. 1987,
Prosser et al. 1993a, 1993b, 1995, Krishnamurthi et al. 1998), data for stars
in IC 2602 (30 Myr) should prove to be particularly useful in trying
to understand the evolutionary progression of stellar rotation from T Tauri
stars (few Myr) to the 50 and 70 Myr ages of Alpha Per and the Pleiades 
respectively. Moreover, 
because of the speed with which rotation changes on the pre-main 
sequence, the differences between these clusters might turn out to be 
important.

Although $v \sin i$ observations exist for IC 2602 (Stauffer et al. 1997), 
these measurements are affected by the unknown angle of inclination, $i$,
and the uncertainties in determining stellar radii. Rotation periods are 
not affected by these uncertainties, and thus they provide more useful 
rotational information. However, greater effort is involved in determining
rotation periods than in obtaining $v \sin i$ measures, a fact which
generally limits the number of period derivations obtained for stars.
Typically, only one half to three quarters of the stars with measured 
$v \sin i$ values yield a rotational period. 
On the positive side, an additional advantage of measuring rotation periods
is that it is possible, if one has the appropriate time coverage, to measure
them even for stars with only $v \sin i$ upper limits.

One of the most puzzling issues to understand in connection with the rotation
rates of stars in young open clusters is the question of the simultaneous 
existence of both ultra-fast and slow rotators in the same clusters. 
Although some data exist for the similarly young cluster IC 2391 (Patten and
Simon 1996), they are by no means an adequate sample. In particular, because 
of the typical duration of observing runs (usually a week or two),
it has not been possible to answer the question of how slow the slowest 
rotators in clusters of this age are. The answer to this question is
of great importance in understanding the extent of the effect of disk 
regulation\footnote{Disk regulation describes the coupling, via magnetic 
fields, of the central star (and hence its rotation) to the young 
circumstellar disk. Edwards et al. (1993) presented good evidence that 
disk-regulation explained the difference in rotation periods between Classical 
and Naked T Tauri stars and this has become the widely accepted paradigm.
If disks do indeed regulate the rotation rates of some stars, a signature of
this regulation should be detectable in the slow rotation rates of older open
cluster stars as well.}
of stellar rotation on the Hayashi track. 
The Ultra-Fast Rotators (henceforth UFRs), on
the other hand, tell us about the necessity and extent of magnetic saturation
in these stars. Thus it is vitally important to obtain a large, unbiased 
sample of rotation information for these stars. In this regard, we have
undertaken a survey (of the rotational properties) of the low mass membership 
of the IC 2602 open cluster in an effort to delineate as much as possible the 
rotational velocity distribution ranging from the UFRs to the slowly rotating 
members having only upper limits on $v \sin i$.

 \section{Observations and Data Analysis}

Because of its far southern location,
IC 2602 has received less attention than it might otherwise have. 
Its  proximity  enables us to monitor its late-type stars with a small 
(1m class) telescope. On the negative side, this proximity results in a large 
apparent cluster size of $\sim$ 100 arcmin (Lynga 1985), and therefore a monitoring 
program cannot be designed to accommodate many program stars on a single exposure. 

In order to avoid unnecessary biases in the selection of candidates to monitor,
we have elected to monitor every previously spectroscopically confirmed 
cluster member from Stauffer et al. (1997) with 
$B-V >$ 0.5. It is probably worthwhile to add that the membership 
assessments in Stauffer et al. (1997) for each of the program stars were made 
after careful consideration of the X-ray emission, photometry, position in the 
color-color diagram, spectroscopy, and where available, lithium abundance.
The reader is referred to that publication (and references therein) for 
further details regarding the cluster membership criteria.

CCD images were taken in the V band, of 23 fields in IC 2602 with the 0.9m 
telescope at CTIO in 1995 March, April and May. They were monitored about once 
a night for a week (1995 March 18-25). Then, after a gap of about 
a week, they were observed more intensively for about 3 weeks (1995 April 4-24)
 with a frequency dictated by weather and other 
observing project constraints, but nominally at the rate of 3-4 times per
night. In addition, a subset of them were monitored, some intensively, by 
CP during 1995 May 8-21. 
The data taken by CP were binned 2x2 pixels while the other 
data were unbinned with a resolution of 0.4 arcsec per pixel. The total
number of sampling points for each star varied, depending on the intensity of
the observing effort, from 42 images for R68 to 155 images for R88A. As far
as observing constraints allowed, we have tried to sample the light
curves of individual stars over the course of the run in a manner that 
would minimize aliasing at 1d intervals. However, as the light
curve for R32 (period $\sim$ 4d) demonstrates, it has not always been
possible to sample a star over all phases.

After CCD processing, including bias subtraction and flat-fielding, using
{\sc iraf}, instrumental magnitudes were measured by fitting 
Point Spread Functions (hereafter PSFs) to the 
stars on each frame using {\sc daophot ii} and {\sc allstar ii} (Stetson, Davis
and Crabtree 1991). Stars on or sufficiently near bad columns were ignored. 
The photometry from the individual frames were converted into time series
using the {\sc daomatch/daomaster} routines from Stetson (1992). These routines
match stars appearing in multiple images according to user-supplied 
specifications, calculate the photometric offsets between frames and output
the corrected time series. The uncertainty in the frame-to-frame magnitude
offsets was less than 0.003 magnitudes for the frames that were eventually 
used because of the large number of stars matched from frame to frame.
The PSF fitting procedure resulted in individual star magnitudes that had 
formal errors of $\sim$ 0.01 mag. for the unbinned data, 
and $\sim$ 0.02 mag. for the binned data. 

This procedure, although CPU intensive, obviates the necessity of choosing a 
few comparison stars for each program star and checking each of them in turn 
for variability or for other problems. The virtue of PSF fitting is that one
need not worry about crowding, and cosmic rays are handled a little more 
gracefully than in aperture photometry. 
A negative is that on nights with particularly bad seeing, we
may have blends of nearby stars. We have usually chosen to discard data
acquired under such conditions. 
The use of an ensemble of stars instead of a few hand-picked ones for 
reference propels the study of stellar variability from the `retail' to the 
`wholesale' arena.  It also allows us 
to check any other objects in the image for variability, if we so choose.  
Indeed, this technique has been employed to discover 
approximately 50 variables in a field centered on the open cluster
NGC 3680 (Platais and Barnes 1998). 

Three techniques
were used to search for periodicity in the IC 2602 target stars - the phase 
dispersion minimization technique of Stellingwerf (1978), the periodogram
technique of Scargle (1982) and the {\sc clean} 
algorithm of Roberts, Lehar and Dreher (1987). 
We have usually run 100 {\sc clean}s with a gain of 0.1 and have searched for
periodicity in the nominal frequency range 0-10 per day. We do not expect any 
of the program stars to rotate any faster than 10 times a day (Barnes and Sofia
1996). Among low-mass stars, the shortest period known so far is 4.4 hrs for
AP124 in the Alpha Per cluster (Prosser et al. 1993b). As for the 
low frequency end, since our sampling extends to over 60 days in some cases,
we expect to be able to detect up to a monthly variability in those fields.
For the most sparsely sampled fields, we are probably sensitive to periods up 
to about 15 days.

For almost all stars whose light curves are provided in this work, the 
variability was observed in the output of all three techniques. In a few 
cases, {\sc clean} was able to pick out a preferred signal where 
none could be found using the techniques of either Stellingwerf or of
Scargle. Once a signal is detected, the evaluation of whether the 
period is believable or not must be made by considering the strength
of the spike in the power spectrum, the variability apparent in the raw
light curve, and the appearance of the phased light curve. For most of
the stars with periodic variations, the variability was happily obvious from 
the raw light curve and the spectral analysis only served to determine the 
period precisely.

\section{Results}

We have monitored 33 late-type stars in IC 2602 with the intent to derive 
rotation periods for these stars by observing photometric variability due to 
modulation of star light by spots. We have detected periodicity in 29
of these stars, of which 26 may be considered firm measurements, and 3 
provisional. 

We display Color-Magnitude Diagrams (CMDs) of the monitored stars in 
Fig. 1\footnote {In this figure and in several following ones, we have chosen 
to mark the stellar positions in the graph with the names of the stars 
themselves for ease of comparison across graphs.}. 
We do not believe that these data are biased with respect to criteria
related to rotation as we have monitored every single star that was then known 
to be a cluster member, independent of its $v \sin i$ value. However, it 
should be kept in mind that the eventual source of the membership is an X-ray 
study (Randich et al. 1995). Whether or not that introduces a bias, we cannot 
at present be sure; but that issue is likely to be resolved soon when a proper 
motion membership survey of this cluster has been completed (Platais 1998). 

Table 1 summarizes the data for each of the stars monitored in IC 2602. 
The first four columns identify the star in question and display the 
photometry from Prosser et al. (1996) and Randich et al. (1995). The fifth 
gives the $v \sin i$ measurement in km/s from Stauffer et al. (1997). The 
sixth gives the number of observations obtained for each star and, in 
parentheses, the time baseline representing the difference in HJD between the 
first and last observations, the seventh and
eighth columns respectively give our period (with the error) and the previous 
period, if any, from Patten et al. (1996).
The ninth displays the Rossby number and the tenth lists 
the $Log (L_x/L_{bol)}$ value from Randich et al. (1995). The eleventh 
displays the amplitude of variability for the monitored stars and the last 
column lists the lithium abundances derived by Randich et al. (1997), where 
available.

  \subsection{Power Spectra and phased Light Curves}

Fig. 2 displays the {\sc clean}ed power spectra for the 29 stars in which 
variability was detected. 
For all stars we plot the power (on an arbitrary scale) as a function
of frequency over the range of 0 to 5 cycles per day.
The spectra 
for those stars with provisional period determinations are marked as such.
In most cases, the spectra are unambiguous in their determination. In the cases
of R24a, R29 and R44, the reason for the secondary spike is obvious in the
phased light curves which are displayed in a corresponding manner in 
Fig. 3. We have displayed two phases along the x-axes. The y-axes, which 
represent the instrumental magnitudes, have arbitrary scales, but we have 
displayed a 0.01 mag. scale so that the reader might have some idea regarding
the relation between the amplitude of variability and the formal photometric
errors.

The light curves (and power spectra) are arranged in order of increasing 
V magnitude (left to right and top to bottom), 
which is approximately the same order as that of increasing color. 
The reader might notice that the bluer stars have somewhat noisier
light curves. In general, the redder ones are cleaner despite their faintness.
This is probably because spots are larger relative to the surface area,
leading to greater modulation of the star light in lower mass stars. 
This issue will be revisited in a
following subsection that deals with variability of the monitored stars.
The structures or features evident in several of the light curves should be noted. 
These features may be interpreted as arising from the details of the physical
structure of the spot patterns on the stars in question. In particular cases, 
over the observing period, we have noticed changes in amplitude (i.e. R72) and
sometimes in period. Additional comments and details about individual stars 
may be found in the appendix.

  \subsection{Rotation Periods}

We have determined rotation periods for 29 stars (3 of which are provisional)
out of the 33 that were monitored. This is a very high success rate and 
attests to the efficacy of our technique. The high success rate for period 
detection may also be traced to the relative youth (and hence high activity) 
of the cluster stars. The phased light curves show clearly that almost all the 
stars are heavily spotted. The periods detected range from about 0.2d (R88A) 
to 10.1d (R77). We believe that these data do not suffer from any selection 
effects related to the length of the observing run.  Although the longest 
period discovered among the stars monitored is only about 10d, the length of 
our observing program ensures that we are able to detect at least 15d periods 
comfortably in all the observed fields, and upto 30d periods in several 
cases\footnote {In fact, we have discovered a variable in the field of R88A 
that has a period of about 43d.}.

A few words about period sensitivity are in order here. While it is clear
that one cannot detect a period longer than the baseline of observations
(in this case, 64d), deciding on the longest detectable period for unevenly
spaced data is a matter of opinion. If the light modulation is large enough
and repeats perfectly and the sampling is good enough, one might expect to
be able to detect even 60d periods, since the overlap on the last 4 days
would be obvious. In practice, light modulation from spotted stars is not
perfectly repeatable because spots evolve and sampling is always non-uniform.
If the investigators are conservative they might demand, as we have, coverage
over two complete cycles but clearly, longer periods, upto the baseline of
the observations, are detectable with good sampling.

Of the four remaining stars without periods, we suspect that R38 and R68 have
rotation periods of 0.79d and 0.66d, compatible with the $v \sin i$
measurements of 48 km/s. We do not trust these 
rotation periods enough to include them in any of the further analysis. The 
remaining two, R21 and R85, are the bluest stars in the sample. Our failure to 
detect variability for these two stars might simply be an indication that they 
do not have large enough spots and therefore would have lower variability
amplitudes.  Their $v \sin i$ values of 23 and 45 km/s 
respectively are certainly large enough that one might expect variability were 
they redder stars.

	\subsubsection{Comparison with $v \sin i$ measurements}

Fig. 4a displays our rotation period measurements against the $v \sin i$
measurements of the same stars by Stauffer et al. (1997). The arrows 
indicate $v \sin i$ upper limits for three stars, R26, R57 and R77, for which 
we have now been able to derive periods. 
The trend of shortening period with increasing $v \sin i$ is, as expected,
clear and no object is located in the top right hand corner.
This information, of course,  makes it possible in principle to calculate the 
angles of inclination of individual stars. We refrain from 
doing so because the errors are large, when we consider the uncertainties 
in the values of the radii of individual stars, along with uncertainties
in the $v \sin i$/period measures. However, it is possible to identify by 
inspection, stars with small and large angles of inclination, and to examine 
whether the light curves of these stars are indeed consistent. 

Stars located along the lower envelope of the points plotted in Fig. 4a
have lower than average $v \sin i$ values which indicate either or both of 
1) a small star and 2) small i.
For the larger and hence bluer ($B-V <$ 1.0) of these stars, a low $v \sin i$
value must therefore indicate a smaller inclination angle and this should be
correlated with what is seen in the corresponding power spectrum, light 
curve and amplitude of variability. 
Conversely, $v \sin i$ values along the upper envelope in Fig. 4a indicate 
either or both of 1) a large star and 2) large i, unless they are also 
associated with unusually short rotation periods.
Thus, for the smaller and hence redder of these stars ($B-V >$ 1.2), 
a high  $v \sin i$ value would indicate a large angle of inclination. We
have examined these issues in relation to these data and indeed have found 
no inconsistency.



	\subsubsection{Color dependence of the Rotation Periods}

Perhaps the most interesting question, and the main motivation for this study,
is how the rotation periods of the IC 2602 stars vary as a function of color.
The mass dependence of the rotation period distribution is particularly
important because models of stellar rotation make predictions of what it
might be, and this information may be used to test them.
These data are particularly useful because they give us a snapshot of 
rotation for solar and late-type stars at 30 Myr. At this point in their
evolution, the IC 2602 stars have descended the Hayashi track and have
ceased to interact strongly with their disks, but have not yet experienced 
main sequence angular momentum loss.

The rotation periods we have derived for our sample of IC 2602 stars 
(including the provisional determinations) are plotted in Fig. 4b as a
function of the $B-V$ color. Each data point is, as before, plotted using
the name of the star and is to be considered as lying at the lower left 
corner of the name. 

One of the features of the data in Fig. 4b (which becomes even more obvious
when the IC 2391 data are combined, as in Fig. 8) is the uniformity of rapid 
rotation across the color range under consideration. Data in the slightly 
older open cluster Alpha Per suggested this for the 0.65 $< B-V <$ 1.25 
(approx. 1 $< M/M_{\odot} <$ 0.8) range.
These data, when combined with those for IC 2391 (Patten and Simon 1996), 
extend this observed behavior over the range 0.5 $< B-V <$ 1.6
(approx. 1.2 $< M/M_{\odot} <$ 0.6). This feature 
of the observations is related to magnetic saturation (e.g. Barnes and Sofia 
1996) and will be discussed in section 4, as will another obvious feature of
these data - the greater dispersion in rotation period among the lower mass
stars.

A relatively surprising feature of these data is the length of the rotation
periods of the slow rotators, which extend to about 10d. Although there are
five stars in the Pleiades ($\simeq$ 70 Myr old) with 7-8d periods, and one 
with an approximately 10d period, such slow rotation was not expected in a 
cluster that is only half as old. Rotation periods among T Tauri stars are 
known to extend to about 16d (with the tail-end of the distribution extending 
to even longer periods) but these stars are expected to spin up significantly 
during pre-main sequence evolution. The slowest rotator known in IC 2391, 
which is also only about 30Myr old, and is generally considered as a twin to 
IC 2602, has a period less than 6d (Patten and Simon 1996). In IC 2602, 
however, 6 stars, representing about a fifth of our sample, have periods 
longer than 6d. The authors opine that small samples and short observing runs 
have conspired to prevent this fact from being discovered earlier. It is 
important to have a long-enough observing run to ferret out the slow rotators. 
If theoretical models (eg. Barnes, Sofia and Pinsonneault 1998, Barnes 1998) 
are to be believed, the existence of these slow rotators cannot be understood 
simply by evolving the slowest T Tauri stars forward in time, but requires the 
additional slowdown provided by a star-disk interaction on the pre-main
sequence.

We can, in fact, use these data to make a somewhat stronger point than the 
one made in the foregoing paragraph. The two slowest rotators, R57 (8.7d) and 
R77 (10.1d), are both quite red, with $B - V$ of 1.60 and 1.47 respectively.
The trend of increasing period with reddening color is the dominant one in the 
Hyades data (Radick et al. 1987) but is not obvious in measurements of younger 
open clusters. Some of this may be ascribed to small sample sizes but much of
it is probably related to the youthfulness of these clusters relative to the
Hyades and to difficulties associated with deriving rotation periods for very
slow rotators. The IC 2602 sample is also a small one, but the fact that the 
slowest rotators are also among the reddest stars in the sample is probably 
worth noting.

In summary, the rotation period data in IC2602/IC2391 indicate that the 
ultra-fast rotators are distributed across the range 0.5 $< B-V <$ 1.6.
We have also derived periods for several slow rotators and note that the 
slowest rotators have redder colors.

  \subsection{Variability Amplitude}

The amplitude of variability (in V mag.) of these stars is plotted in 
Fig. 5a as a function of the $B-V$ color. 
Contrary to expectations that earlier type stars, because they are usually 
not as heavily spotted as their lower mass counterparts, might have smaller 
variability amplitudes, no strong dependence is immediately obvious from 
these data. This could be an indication that all of the IC 2602 (and IC 2391)
stars are so young and hence still so active that the differences between
the earlier and later type stars is not yet obvious. This viewpoint seems to 
be supported by the X-ray data, which is discussed in the following section.

There is, however, a weak trend in these data that merits attention. 
If we divide Fig. 5a into four quadrants - blueward and redward of 
$B-V =$ 1.0 and variability amplitudes less than and greater than 0.04 
mag., only R58 appears in the second quadrant. This star is peculiar 
because in SB's data set, its variability amplitude was only 0.02, while in 
CP's data it has the much larger variation that is plotted in Fig. 5a. 
Regardless of the position of R58, one can make the following statement: 
The majority of the blue stars have variability amplitudes less than 0.04 
mag., while the majority of the red stars have amplitudes in excess of 
this. The conclusion seems to be bolstered slightly by the additional data 
available in IC 2391 (Patten and Simon, 1996) which is plotted in Fig. 5
using open circles.
It will be interesting to examine whether it will continue to hold when
larger data sets become available. 

The above viewpoint receives support in Fig. 5b, which displays the 
amplitude of variability as a function of rotation period. This figure clearly 
separates the IC 2602 stars into two distinct groups. The ones with
periods shorter than 6d show a wide range in the amplitude of variability, 
from 0.01 to 0.08 V mag.
Those with longer periods all have amplitudes less than or equal to only 0.03 
mag. This is highly relevant to period determinations in older clusters,
because it suggests that it will be difficult to detect variability in similar
stars when they are older and their activity has decreased. 

Interestingly, all the long period stars (with small variability amplitudes) 
are later types with $B-V >$ 1.0, and this despite the fact that 
late-type stars have greater amplitudes of variability on average than early 
type ones. Also, among the stars with periods shorter than 4d, every single 
star (with the exception of R58) that has an amplitude greater than 0.04 
mag. is a red star. The blue stars are not as variable as we saw above. 
This suggests that if a blue star were a slow rotator,
it would probably go undetected in the present period searches. 
Our sample does not contain such a star. 

The information contained in the foregoing two figures can be 
combined together by displaying the variability amplitude as a function
of Rossby number\footnote {The Rossby number
is defined as $P_{rot}/\tau_c$, where $P_{rot}$ and $\tau_c$ represent the 
(observed) rotation period and the (calculated) convective turnover timescale, 
respectively.}. For completeness, we have displayed this information for our 
sample of stars in IC 2602 in Fig. 5c. 
From the viewpoint of the Rossby number calculation, the rotation 
period dependence appears directly in the numerator, and the color dependence 
of the variability appears in the denominator via the convective turnover 
timescale, which is an equivalent variable for these purposes. 
We have used the convective turnover timescales provided by Kim and 
Demarque (1996). 

  \subsection{X-rays}

The connection between rotation and X-ray emission is particularly interesting
in view of the concept of magnetic saturation, which suggests that a 
stellar magnetic field does not increase indefinitely with a star's angular
velocity, but saturates beyond a threshold value that varies from star to 
star in a manner not yet understood\footnote {In the traditional picture,
$\frac{dJ}{dt} \propto \Omega^3$, and $\Omega \propto B$, where 
$\frac{dJ}{dt}$, $\Omega$ and $B$ represent the angular momentum loss rate, the
rotation rate and the magnetic field of the star, but if this were always 
true, we would not have rapid rotation.}.

Some intriguing observational evidence in support of magnetic saturation
was provided by Stauffer (1994) and Stauffer et al. (1994), 
who noticed that the X-ray emission (which
is widely regarded as an indicator of magnetic field strength) of individual
stars in the Pleiades has a wide range for stars with low $v \sin i$ values,
but stays relatively constant at an elevated level for those with high values.
Models that included saturation were calculated by Chaboyer et al. (1997) and
others but the results were ambiguous because the models did not start high
enough on the Hayashi track and plentiful observations were not available then.
The idea received a further boost when it was realized (Barnes and Sofia 1996; 
Krishnamurthi et al. 1997) that theoretical models of rotating stars that 
started at the stellar birthline also could not explain the existence of
the ultra-fast rotators in young star clusters without incorporating the 
idea of magnetic saturation (with a threshold that varied with stellar mass).

Given the extensive X-ray coverage of this cluster, and the considerable 
interest in stellar activity, it might be worthwhile to examine how these 
data relate to the X-ray properties of this cluster. The X-ray data 
(Randich et al. 1996) have played a crucial part in the study of this 
cluster by taking over the role usually played by a proper motion study. 
The stars identified as candidate members of the cluster by the X-ray data 
have been studied photometrically (Prosser et al. 1996), and spectroscopically 
(Stauffer et al. 1997), to a point that we are now reasonably certain about 
the cluster membership.

This raises the issue of whether the IC 2602 members also show signs of 
magnetic saturation. Fig. 6a displays the distribution of $Log (L_x/L_{bol})$ 
as a function of the observed rotation periods from Table 1. It is clear
that many of the stars are operating in the saturated regime.
The scatter in $L_x/L_{bol}$ seen
in these diagrams may be attributable to various influences. For instance,
previous studies of X-ray activity in other relatively young clusters
(eg. Randich et al. 1996, Figs. 9/10) indicate that some of the scatter
in $L_x/L_{bol}$ seen for IC 2602 in Fig. 6 is due to the range in spectral
class covered. In the present study, R21, R83 and R85 are among the
earliest spectral range of stars from Table 1 shown in Fig. 6. R70 (=W85) was 
also listed as having a relatively early spectral type (F7, Whiteoak 1961)
which would concur with the relatively low level of X-ray activity seen.
In addition, in some rare instances the X-ray flux measurements for certain
stars may have been slightly underestimated, resulting in an abnormally low
$L_x/L_{bol}$ value. Underestimates of the X-ray flux would occur due to
instrumental effects as explicitly discussed in connection with the raster
X-ray survey in Alpha Persei (Randich et al. 1996, sect. 3.2.1), but which were
not addressed during the earlier analysis of the raster X-ray survey in
IC 2602 (Randich et al. 1995). Possibly more evident among slow rotators,
some stars with slightly higher than usual X-ray activity levels may have
been observed during a state of high X-ray emission, such as a flare or
post-flare state, resulting in overestimates of their normal quiescent X-ray
flux levels. Overall, the distribution in X-ray activity as a function of
rotation for IC 2602 members compares favorably with similar distributions
constructed for other clusters such as the Pleiades and Alpha Persei; stars
with $v \sin i > 15$ km/s or P $\le$ 2d exhibiting X-ray activity near or
at the saturation level.

The question of magnetic saturation is further complicated by a color 
dependence of the observed X-ray activity levels. In Fig. 6b
the distribution is illustrated as a function of Rossby number which 
yields a somewhat clearer trend with X-ray activity than just the period or 
color dependence by itself, presumably reflecting the fact that X-ray activity
is a function of both. The distribution in Fig. 6b is substantially similar to 
that in Stauffer et al. (1997, Fig. 7), but we have used the true rotation 
periods of the IC 2602 stars rather than the $v \sin i$ observations in 
calculating the Rossby number. 

One final point deserves to be mentioned: the amplitude of variability does
not seem to show any identifiable trend just yet with the X-ray emission.
Such a trend might be expected since, as we have shown in Fig. 5, there
are trends with color and period, and hence with Rossby Number. 
This probably indicates either that this
cluster is not old enough for the trend to be obvious, or is a result of 
having an insufficiently large sample. The trend might emerge when 
the current database has been expanded substantially.

  \subsection{Lithium}

Butler et al. (1987) first suggested a connection between Lithium abundance
and rotation, showing that four rapidly rotating early-K dwarfs in the 
Pleiades contained an order of magnitude more lithium than four slow rotators 
of the same spectral type. 
Observations by Soderblom et al. (1993) showed unambiguously (using $v \sin i$
data) that fast rotators in the Pleiades had higher than average lithium 
abundances than the slow rotators. 
Jones et al. (1997) show a similar trend in M34 (again using $v \sin i$ data). 
Thus, it might be worthwhile to ask about trends in IC 2602.

Randich et al. (1997) have determined lithium abundances for many of the 
IC 2602 stars. They show that the overall trends of lithium abundance with 
temperature are similar to those in the Pleiades and Alpha Per. In addition, 
they compare the lithium abundances to the $v \sin i$ observations available 
(Fig. 12 of Randich et al.). In that figure, the two stars with $v \sin i$ 
greater than 30 km/s do indeed lie above the trend for the slower rotators. 
We would like to remove the ambiguities associated with $v \sin i$ data by 
replacing them with rotation periods. Fig. 7 displays the LTE lithium 
abundances from Randich et al. (1997) against the $B-V$ color, marking the 
stars using their rotation periods (stars without period determinations are 
marked 99.99).  Despite the paucity of the data points (there are several
stars with periods but no lithium abundances and vice versa), it is 
interesting that, indeed, the short period rotators are located above the 
mean cluster trend. It might be worthwhile to determine abundances for the 
remaining unmeasured stars in this cluster and also for those in IC 2391.

\section{Interpretation using stellar models}

One of the goals of the modern study of stellar evolution is to understand 
the rotational
evolution of stars. As soon as new data for individual clusters become 
available, they offer us a chance to compare the new observations with the
current paradigm which may then need to be altered and perhaps even discarded.
In the case of IC 2602, we have not found any new data that would necessitate
discarding the current paradigm, but the new data appear to clarify several
issues and modify one important one. The major results of these data appear
to be the delineation of the ultra-fast rotators across the 0.6 $< B-V <$ 1.6
color range, and the identification of several very slow rotators.

The main problem in stellar rotation is to achieve an understanding of 
the evolution of stellar rotation with age. The important issues concern the 
internal transport of angular
momentum with age, the effects of magnetic saturation, and those of
disk-interaction. This last effect has assumed an added importance with
the discovery of extra-solar planets. A star-disk interaction on the 
pre-main sequence has been suggested (eg. Edwards et al. 1993) as the reason 
for the dichotomy in the observed rotation periods of
Classical and Naked T Tauri Stars. If this were the case, the fast end of
the T Tauri distribution should evolve into the fast end of the distributions
for young open clusters and the slower T Tauri stars should evolve into the
slower rotators in young clusters, with reasonable disk-interaction timescales.
Alternatively, one could reverse the argument and use the observed slow
rotators in young open clusters to derive disk-interaction timescales.

We will not discuss details of the various theoretical models here, preferring
to present them elsewhere (Barnes et al. 1998, Barnes 1998), but we will
compare one set of models to the observations presented here. 
The rotation periods of the IC 2602 stars from Table 1 and those of their 
twin cluster IC 2391 (Patten and Simon, 1996) are plotted in Fig. 8 against 
$B-V$ color. We have also overplotted a number of `isoarchs'
taken from Barnes (1998) each of which 
represents the locus of points with different stellar mass but the same 
initial conditions. The lowest of these corresponds to stellar models of 
masses 1.2, 1.0, 0.8 and 0.6 $M_\odot$ that were started off with 4d initial 
periods, and had no disk interaction. The next one consists of the same models,
 but with a 16d initial period. The ones above all have initial periods of 16d,
 but have been disk-locked for successively longer times ranging from 0.3 Myr
to 10 Myr. 
If the models are correct, then the observations of the ultra-fast rotators
in these clusters are consistent with 4d initial periods and no disk-locking
whatsoever.

The behavior of the slow rotators is more surprising. Most of them lie 
above (i.e. they are slower than) the isoarch with 16d starting period and 
no disk-locking. 
Thus, the models predict a very narrow range of rotation periods for stars 
without disks regardless of the initial period. 
Since the majority of the observations lie above the isoarch for a 16d initial
period and no disk locking, the conclusion seems to be inescapable that the 
majority of stars in these open clusters, and perhaps all open clusters, have 
been disk-locked for significant periods of time. 
Although one of the slow rotators is the problematic star B134, 
(which has been designated as a photometric binary), 
there are some other very slow rotators and the models 
seem to require disk-lifetimes up to 10 Myr to explain them. The majority, at
least in the 1.2 to 0.6 $M/M_{\odot}$ range under consideration here, seem 
to be explainable with disk-lifetimes up to 5 Myr.
With these data alone, we cannot say whether any particular disk-lifetime
is preferred as there does not seem to be a concentration about any particular
isoarch. 

\section{Conclusions}

We have monitored 33 late-type stars in the 30 Myr old open cluster IC 2602 
for photometric variability due to spot modulation. 
Rotation periods were derived for 29 of these stars (including 3 for which
only $v \sin i$ upper limits existed before). 
The rotation periods derived range from 0.2d to 10d, 
representing the entire range from the ultra-fast rotators to the slowest 
rotators yet discovered in a cluster of this age.  

We find rapid rotation across the entire 0.5 $< B-V <$ 1.6 color range. 
The existence of rapid rotation may be interpreted as evidence of magnetic
saturation, since theoretical models of rotating stars cannot create 
ultra-fast rotators otherwise. While the IC 2602 sample here covers a
somewhat limited mass range, the uniformity of the color dependence 
may possibly be interpreted as evidence for a mass-dependent saturation 
threshold. 

The range of rotation periods observed at any color in this cluster is so
wide that theoretical models of angular momentum loss are unable to explain
their existence without some additional effect, such as a star-disk 
interaction leading to rotational slowdown. If the theoretical models are 
correct, then the majority of the IC 2391 and IC 2602 stars have experienced 
significant disk-locking (of the order of Myrs) on the pre-main sequence. 

The amplitude of variability of individual stars displays trends with both
color and rotation period. Bluer stars have smaller variability amplitudes
as do stars with rotation periods longer than 6d. These behaviors might
help in understanding the selection effects of the presently available
rotation period data and in planning future observations.

Overall, the distribution in X-ray activity as a function of rotation for
IC 2602 members compares favorably with similar distributions constructed
for other clusters such as the Pleiades and Alpha Persei; stars with 
$v \sin i > 15$ km/s or P $\le$ 2d exhibiting X-ray activity near or at
the saturation level. The weak trend between X-ray activity level and Rossby
number previously noted for this and other clusters (Stauffer et al. 1997,
Patten and Simon 1996) continues to hold.

Lastly, we have compared our period determinations with the lithium abundance
determinations of Randich et al. (1997). The small size of the sample precludes
definitive statements but the shorter period stars seem to display elevated 
lithium abundances relative to slower rotators of the same color.

{\it Acknowledgements.} 
Part of this work was completed while SB held the Lowell Fellowship, funded by
the BF Foundation, at Lowell Observatory. SB would also like to thank Robert 
Zinn for his support, Jerry Orosz and Eric Rubenstein for their generous help 
with all aspects of the observation and reduction of these data, and Yasuhiro 
Hashimoto for help with IDL. CFP/JRS acknowledge support for this research 
from NASA grant Nos. NAGW-2698 and NAGW-3690. The assistance and support of 
the CTIO staff and personnel are gratefully acknowledged in contributing to 
the successful undertaking of the observations.

\appendix
\section{Appendix: Comments on individual stars}

{\it \noindent B134}: This star is a photometric binary.\\
{\it W79}: The approximately 6d variability is visible directly in the raw light curve which also suggests some spot evolution.\\
{\it R15}: Provisional period with smallest amplitude of variability. Secondary spike chosen in spectrum. Stellingwerf analysis also gives most power at 3.6d.\\
{\it R21 (B6)}: No suitable comparison stars were available in images, preventing analysis.\\
{\it R24A}: Structure in the phased light curve suggests two major spot groups. Shape of light curve remains the same even if CP's data is ignored.\\
{\it R26}: 5-6d variability obvious from the raw light curve.\\
{\it R27}: Provisional period. Due to possible phase shifting in data over the combined dataset, the plotted light curve shows only SB's data. We suspect two major spot groups on this star. \\
{\it R29}: Messy light-curve. Excising various segments of the data does not reduce the scatter. Variability is apparent from the raw light curve. Period of 2.21d is essentially the same as the previously known period (Patten et al. 1996).\\
{\it R31}: Messy light-curve. Excising various segments of the data reduces the scatter put produces a less convincing phased curve. Half-day variability is apparent from the raw light-curve itself. Amplitude of variability is small.\\
{\it R32}: The 4d period is believable despite being a multiple of a day. Variability is large and unmistakable in the raw light curve. The 4d periodicity results in a light curve with missing phase coverage.\\
{\it R38}: No strong peak is visible in periodogram based on 92 combined observations. Using 46 observations in the Prosser data for this star, CP finds a suggestive period at P $\simeq$ 0.8d = 19 hr, with an amplitude of $\Delta V =$ 0.04 mag.\\
{\it R43}: We confirm the previous period of 0.78d (Patten et al. 1996, Prosser 1997).\\
{\it R44}: The highest power is visible at twice the chosen frequency but the correct period can be picked up by inspection of the raw light curve which shows the 5-6d variability clearly.\\
{\it R50}: Period derived here confirms the previously suspected 6.4d period (Patten et al. 1996).\\
{\it R52}: The previously derived period (Patten et al. 1996) is confirmed.\\
{\it R53b}: Phased light curve shows sub-structure, similar to that displayed by other rapid rotators, possibly because the spot pattern changed during the run.\\
{\it R56}: The approximately 4d periodicity is apparent from the raw light curve and appears in all three parts of the observing run. The phased light curve is a little noisy. Phase change between SB's data and CP's gives the appearance of two superposed light curves. Only SB's data is displayed in Fig. 3.\\
{\it R57}: Significant detection in periodogram, though period error approaches $\sim$ 1d for this star due to more limited coverage for this long period.\\
{\it R58 (B102)}: The same periodicity of $\sim$ 0.57d persists in various subsets of the data. The light curve appearance using all observations is somewhat degraded due to small changes in phase between data subsets. This may be interpreted as evidence of differential rotation or migration of starspots. Amplitude of variability is $\sim$ 0.02 mag. in SB's data, increasing to $\sim$ 0.055 in CP's observations.\\
{\it R66}: Variability is apparent in the raw light curve. However, the behavior of the star changed in the second half of the run, resulting in a good phased light curve once this data is deleted.\\
{\it R68 (W84)}: Strong aliasing effects prevent specific period determination for this star. Based on a subsample of 30 observations and observed $v \sin i$ for R68, CP indicates a suggestive period at P $\simeq$ 0.66d $\sim$ 16 hr, with amplitude $\Delta V =$ 0.04 mag. Previous work (Patten et al. 1996) suggests a provisional 0.99d period.\\ 
{\it R70 (W85)}: The variability is apparent from the light-curve but the phased light curve does not look as good as would be expected, probably because of the small amplitude of variability.\\
{\it R72 (B120)}: The spectrum is double-peaked, likely due to the star's change in behavior during the run. The star is systematically brighter in CP's dataset.\\
{\it R77}: Variation occurs in light curve appearance between SB/CP's datasets; only SB's data is displayed in the phased curve. Formal period error is on order of 1 day.\\
{\it R83}: Phased light curve shows SB's dataset.\\
{\it R85 (B131)}: No period determination. Variation approaching 0.1 mag. seen in CP's data, however.\\
{\it R88A}: This is the fastest rotator. We confirm the previously derived period (Patten et al. 1996, Prosser 1997).\\
{\it R89}: The 4.8d period found here is slightly different from the previously estimated period of 4.47d (Patten et al. 1996, Prosser 1997).\\ 
{\it R92 (B132)}: Provisional period. The raw light curve seems to show some 2d periodicity. Phase coverage is not good because the period is a multiple of a day. Photometric binary?\\
{\it R93}: Some points (not all in the same dataset) are off the sequence on the phased light curve, indicating irregular variations superimposed on the rotational modulation.\\ 
{\it R94}: This star has a nearby companion as indicated in Randich et al. (1995). Available photometry suggests a photometric binary in $V-I$.\\
{\it R95A}: The first week's data lies above the phased sequence so the light curve looks better without it. The previous period from Patten et al. (1996) is confirmed.\\ 
{\it R96}: Periodicity under 2d is apparent from the raw light curve. Previously known to have a similar period (Patten et al. 1996, Prosser 1997).\\ 
 

\newpage
\begin{center}
Table 1: Data summary for IC 2602 stars.
\end{center}
\tiny{
\begin{tabular}{*{11}{c}p{1.0in}}
\hline\\
Name&V&$B-V$&$(V-I)_c$&$v \sin i$&N$_{\rm obs}$(Baseline)&Period&P$_{prev}$&$N_r$&$Log\frac{L_x}{L_{bol}}$&VarAmp&n(Li)\\
    & &     &         &  (km/s)  &     &  (d) & (d)    & &                & (mag)&     \\
\hline\\
B134 & 10.66 & 0.95 & 1.00 & 10 & 61(64d)& 6.8 $\pm0.3$ &...	& 0.110 & -3.69 & 0.025 & ...\\
W79  & 11.57 & 0.83 & 0.85 & 8  & 61(37d)& 6.2 $\pm0.5$ &...	& 0.117 & ...   & 0.02  & ...\\
R15  & 11.75 & 0.93 & 1.06 & 7  & 41(37d)& 3.6:$\pm0.2$ &...	& 0.059 & -2.96 & 0.01  & 2.56\\
R21  &  9.50 & 0.51 & 0.62 & 23 & 37(37d)& ...  	   &...	&  ...  & -3.83 &  ...  & 3.32\\
R24A & 14.61 & 1.43 & 1.86 & 34 & 93(64d)&1.25$\pm0.01$ &...	& 0.013 & -3.07 & 0.06  & ...\\
R26  & 15.14 & 1.54 & 2.15 &$<$6& 93(64d)& 5.7 $\pm0.3$ &...	& 0.056 & -3.57 & 0.05  & ...\\
R27  & 14.35 & 1.50 & 1.80 & 10 & 57(64d)& 4.5:$\pm0.2$ &...	& 0.045 & -3.57 & 0.035 & ...\\
R29  & 12.73 & 1.11 & 1.19 & 22 & 77(64d)&2.21$\pm0.04$ &2.19& 0.031 & -3.21 & 0.05  & 1.76\\
R31  & 15.08 & 1.59 & 2.24 & 35 & 83(64d)&0.49$\pm0.01$ &...	& 0.005 & -2.94 & 0.02  & ...\\
R32  & 15.06 & 1.63 & 2.16 & 9  & 39(51d)& 4.0 $\pm0.2$ &...	& 0.035 & -3.16 & 0.075 & ...\\
R38  & 15.72 & 1.53 & 2.50 & 48 & 92(64d)& ...  	   &...	&  ...  & -3.10 & ...   & ...\\
R43  & 12.14 & 0.95 & 1.10 & 50 & 42(53d)&0.78$\pm0.01$ &0.78& 0.013 & -3.01 & 0.035 & 3.17\\
R44  & 14.88 & 1.55 & 2.03 & 7  & 57(64d)& 5.5 $\pm0.2$ &...	& 0.053 & -3.60 & 0.04  & ...\\
R50  & 14.75 & 1.56 & 2.08 & 7  & 80(64d)& 6.4 $\pm0.4$ &... & 0.061 & -3.19 & 0.02  & ...\\
R52  & 12.19 & 1.07 & 1.11 & 95 & 76(64d)&0.393$\pm0.002$&0.393& 0.006 & -3.33 & 0.07  & ...\\
R53B & 15.39 & 1.61 & 2.49 &100 & 58(53d)&0.41$\pm0.01$ &...	& 0.004 & -3.12 & 0.045 & ...\\
R56  & 13.64 & 1.43 & 1.60 & 17 & 88(64d)& 4.1 $\pm0.1$ &...	& 0.043 & -2.99 & 0.055 & $<$0.65\\
R57  & 15.59 & 1.60 & 2.44 &$<$6& 40(37d)& 8.7 $\pm0.9$ &...	& 0.079 & -3.30 & 0.02  & ...\\
R58  & 10.57 & 0.65 & 0.76 & 93 &108(64d)&0.57$\pm0.01$ &...	& 0.015 & -3.22 & 0.055 & 3.55\\
R66  & 11.07 & 0.68 & 0.83 & 12 & 39(51d)& 3.3 $\pm0.2$ &...	& 0.082 & -3.76 & 0.035 & 3.10\\
R68  & 11.28 & 0.89 & 1.09 & 48 & 42(37d)& ...  	   &0.99:&  ...  & -3.03 & ...   & 2.90\\
R70  & 10.92 & 0.69 & 0.71 & 11 & 63(64d)& 4.3 $\pm0.1$ &...	& 0.107 & -4.44 & 0.02  & 2.95\\
R72  & 10.89 & 0.64 & 0.76 & 49 & 65(63d)&1.05$\pm0.02$&...	& 0.029 & -3.01 & 0.035 & 3.41\\
R77  & 14.12 & 1.47 & 1.72 &$<$7& 70(64d)& 10.1$\pm0.9$ &...	& 0.104 & -3.66 & 0.03  & ...\\
R83  & 10.70 & 0.62 & 0.78 & 30 &135(64d)&1.67$\pm0.01$ &...	& 0.049 & -3.55 & 0.035 & 3.45\\
R85  &  9.87 & 0.52 & 0.58 & 45 &110(64d)& ...  	   &...	&  ...  & -4.74 & ...   & 3.17\\
R88A & 12.71 & 1.20 & 1.35 &200 &135(64d)&0.204$\pm0.001$&0.204& 0.003 & -3.54 & 0.04  & ...\\
R89  & 12.97 & 1.24 & 1.35 & 14 & 51(37d)& 4.8 $\pm0.3$ &4.47& 0.059 & -3.49 & 0.065 & 1.60\\
R92  & 10.26 & 0.67 & 0.78 & 14 & 34(37d)& 2.0:$\pm0.1$ &...& 0.050 & -3.72 & 0.015 & 3.12\\
R93  & 13.79 & 1.37 & 1.62 & 8  & 53(64d)& 6.7 $\pm0.4$ &...	& 0.075 & -3.60 & 0.025 & ...\\
R94  & 13.33 & 1.39 & 1.73 & 23 & 48(64d)& 2.6 $\pm0.1$ &...	& 0.028 & -3.56 & 0.035 & $<$0.87\\
R95A & 11.73 & 0.87 & 0.97 & 12 & 51(51d)&1.20$\pm0.02$ &1.23& 0.021 & -2.90 & 0.04  & 3.23\\
R96  & 12.94 & 1.25 & 1.37 & 17 & 36(37d)&1.82$\pm0.05$ &1.81& 0.022 & -3.18 & 0.05  & ...\\
\hline\\
\end{tabular}
} 

\clearpage
 
\begin{figure}[1]
\plotfiddle{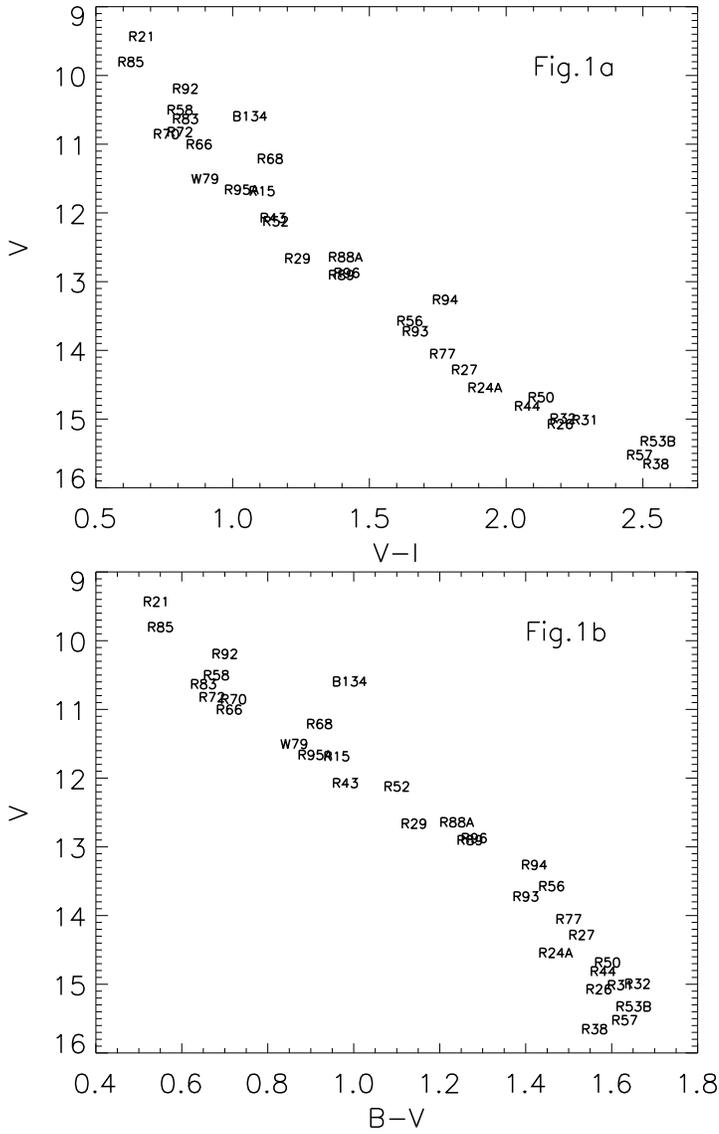}{8in}{0}{100}{100}{-310}{-50}
\caption[fig1.ps]{Color-magnitude diagrams for the monitored stars in IC 2602 for $V-I_{\rm c}$ (upper panel) and $B-V$ (lower panel).}\label{fig1}
\end{figure}
 
\begin{figure}[2]
\plotfiddle{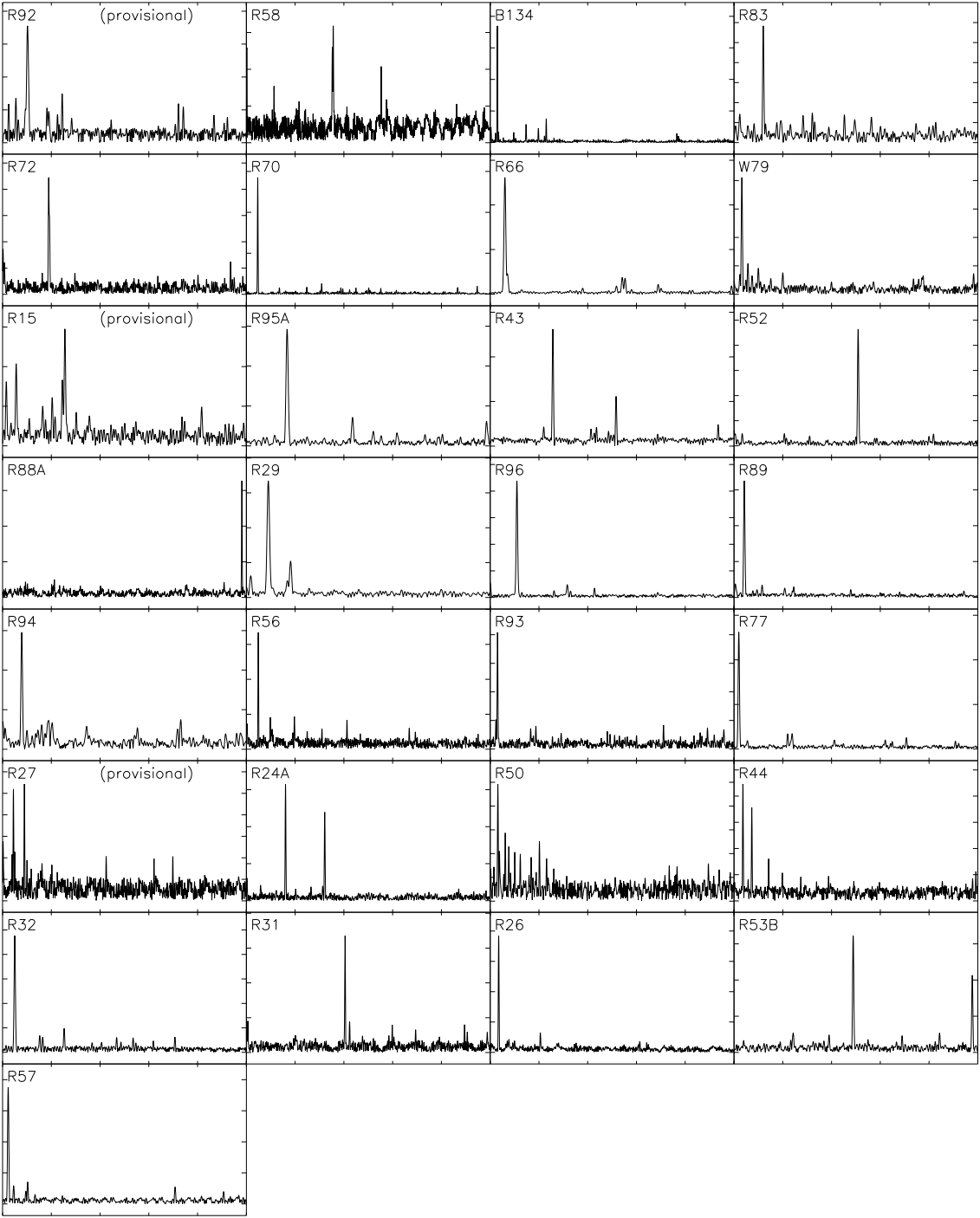}{8in}{0}{80}{80}{-310}{-50}
\caption[fig2.ps]{{\sc clean}ed spectra for the 29 variable stars for which we have determined rotation periods. Stars are arranged in order of increasing V mag., right to left and top to bottom. The frequency (x) scale ranges from 0 to 5 $cycles/day$. The power (y) axes are arbitrary.}\label{fig2}
\end{figure}
 
\begin{figure}[3]
\plotfiddle{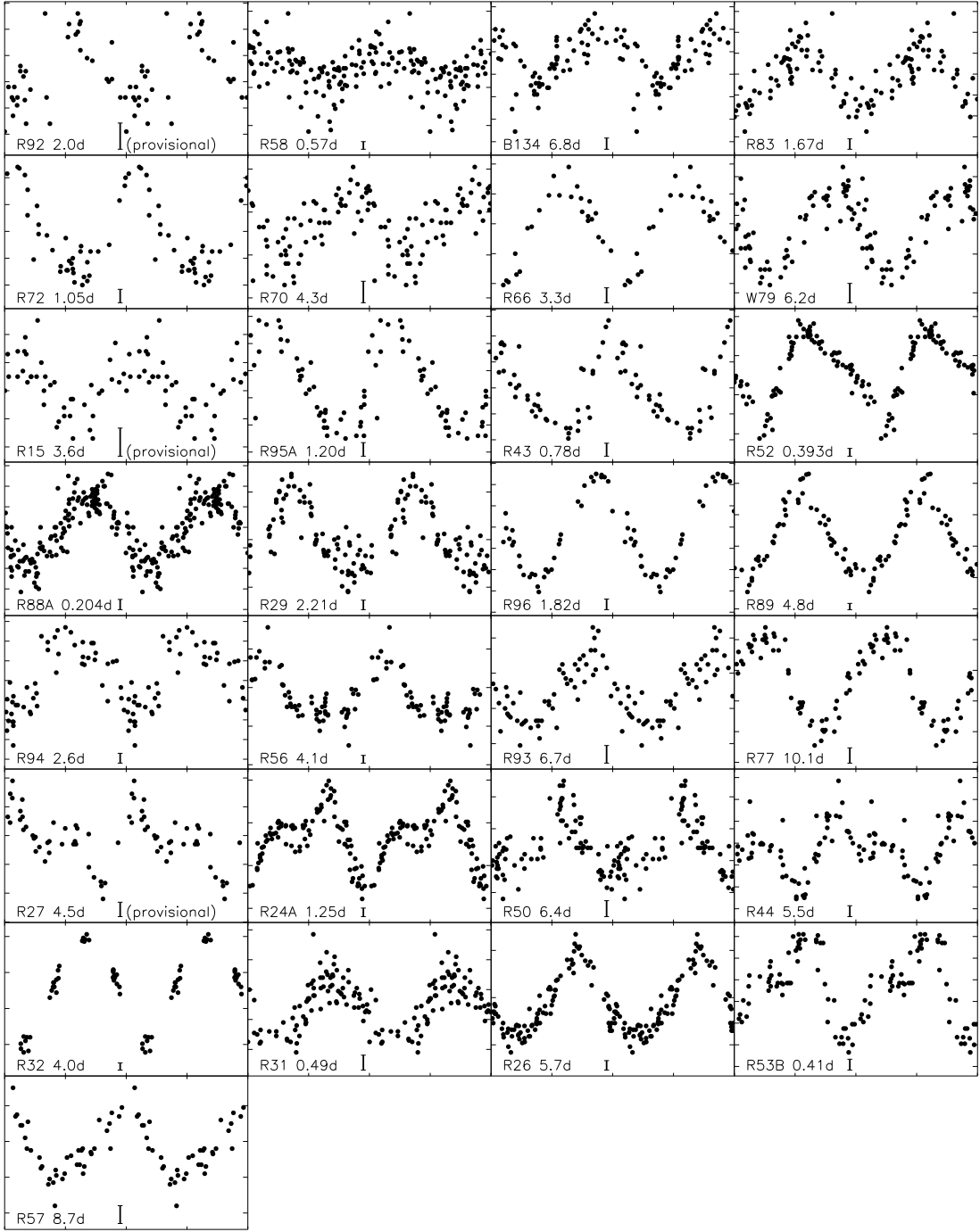}{8in}{0}{80}{80}{-310}{-50}
\caption[fig3.ps]{Phased light curves for stars in IC 2602 for which we have been able to derive rotation periods. Two phase cycles are plotted, brightest points at the top. The magnitude (y) axes are arbitrary but the scale bar shown in each panel corresponds to 0.01 V mag.} \label{fig3}
\end{figure}
 
\begin{figure}[4]
\plotfiddle{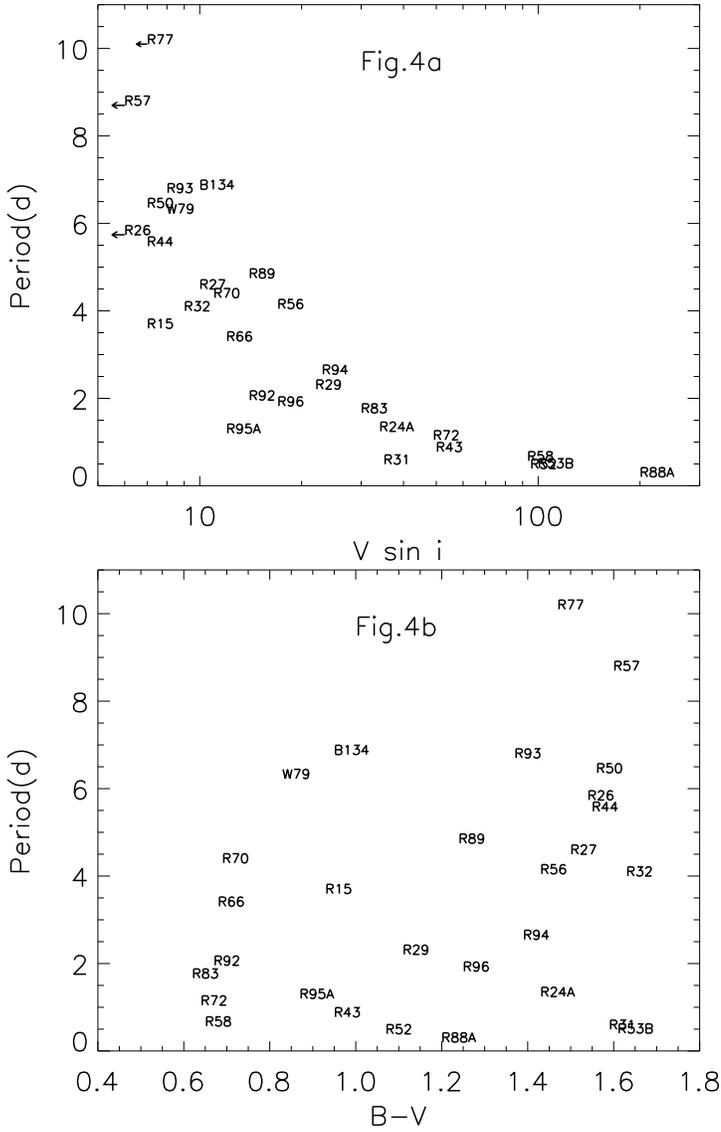}{8in}{0}{100}{100}{-310}{-50}
\caption[fig4.ps]{Rotation periods of IC 2602 stars vs $v \sin i$ (upper panel) and $B-V$ color (lower panel). Arrows indicate upper limits for $v \sin i$. Note that there are several stars with periods longer than 6d and that the two slowest rotators are also among the reddest stars in the sample.}\label{fig4}
\end{figure}

\begin{figure}[5]
\plotfiddle{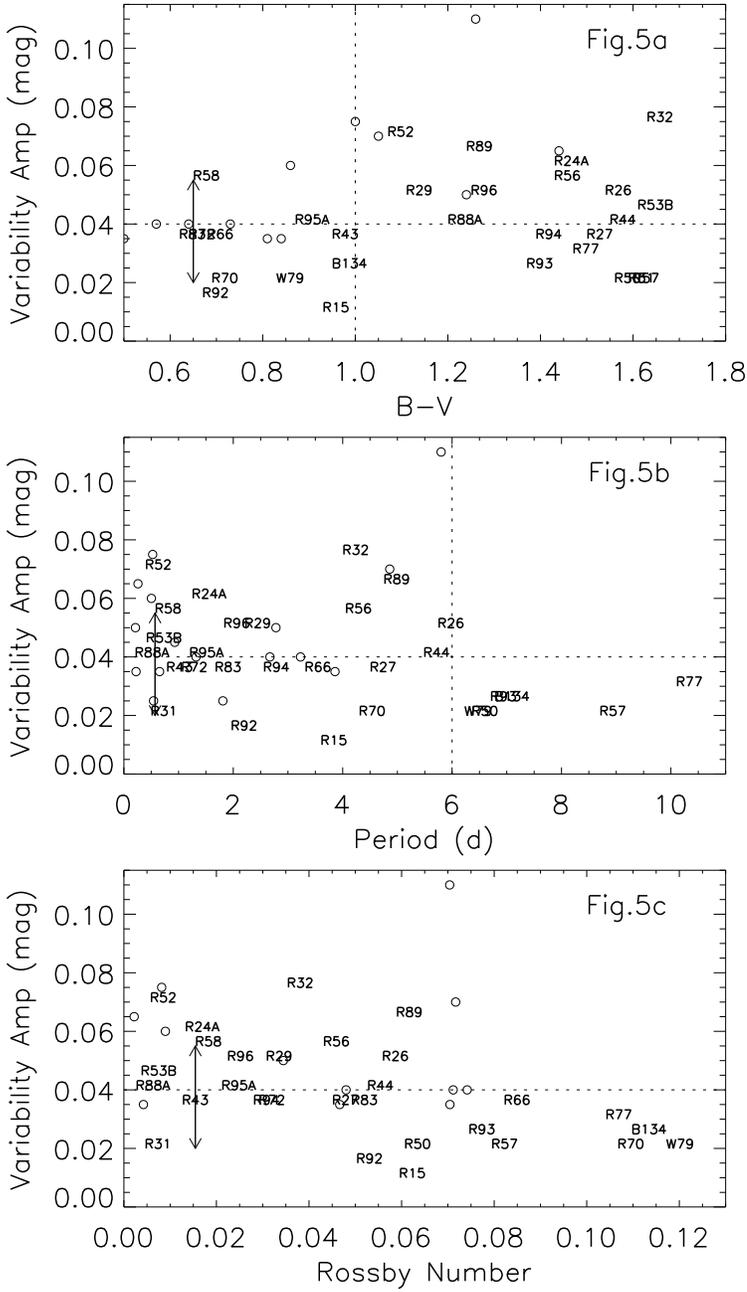}{8in}{0}{100}{100}{-310}{-50}
\caption[fig5.ps]{Amplitude of variability of IC 2602 stars as a function of $B-V$ color (upper panel), period (middle panel) and Rossby number, $N_r$ (lower panel). The arrow for R58 indicates the two different amplitudes in SB's and CP's data. Open circles denote IC 2391 data from Patten et al. (1996). Note the small amplitudes for the blue and the long-period stars.}\label{fig5}
\end{figure}
 
\begin{figure}[6]
\plotfiddle{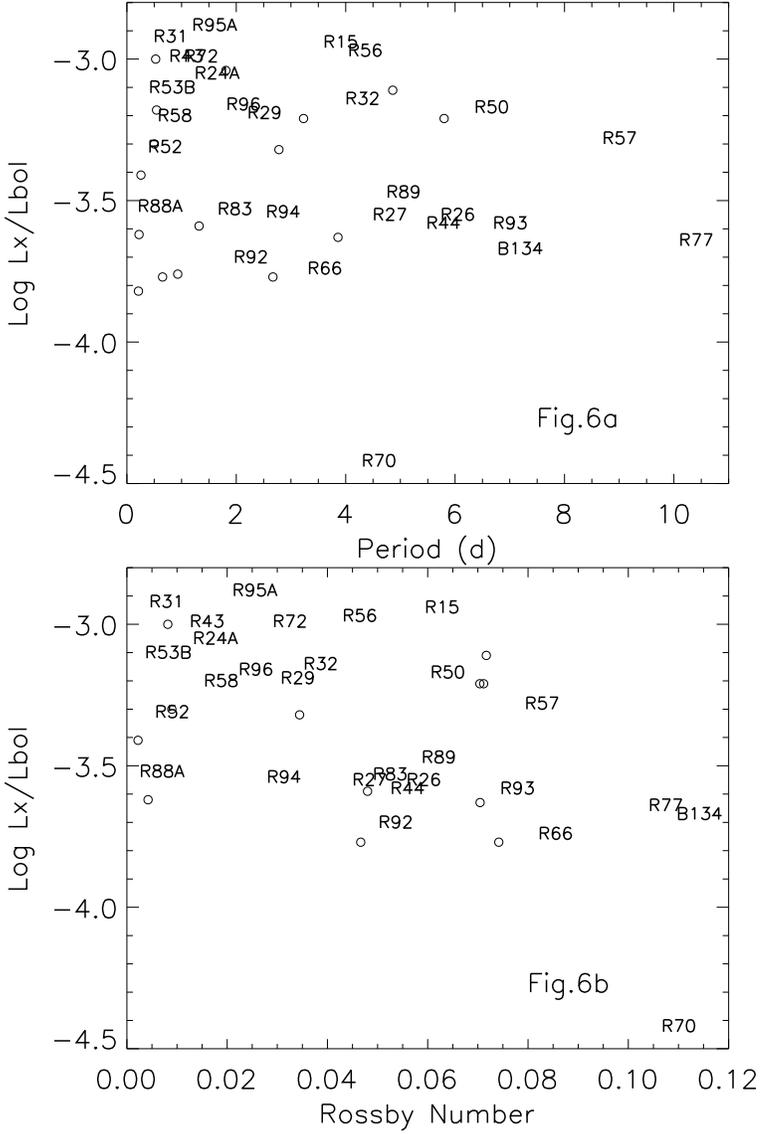}{8in}{0}{100}{100}{-310}{-50}
\caption[fig6.ps]{Xray activity (Randich et al. 1996) of IC 2602 stars vs rotation period (upper panel) and Rossby number, $N_r$. (lower panel). Open circles denote IC 2391 data from Patten \& Simon (1996).}\label{fig6}
\end{figure}

\begin{figure}[7]
\plotfiddle{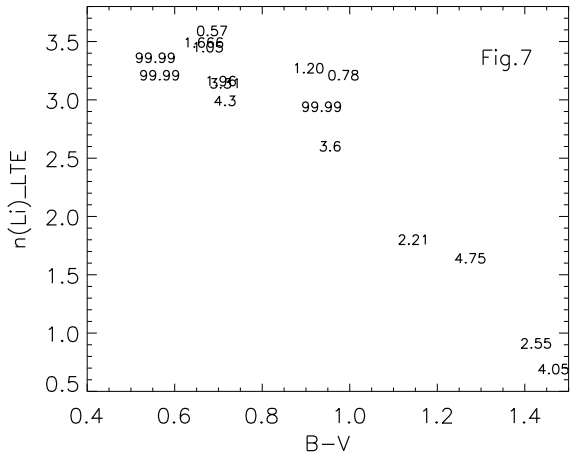}{8in}{0}{80}{80}{-310}{-50}
\caption[fig7.ps]{Lithium abundances of IC 2602 stars from Randich et al. (1997). Stars are marked using their rotation periods (in days) and those without period derivations are marked `99.99'. Note the elevated abundances of the short-period stars.}\label{fig7}
\end{figure}

\begin{figure}[8]
\plotfiddle{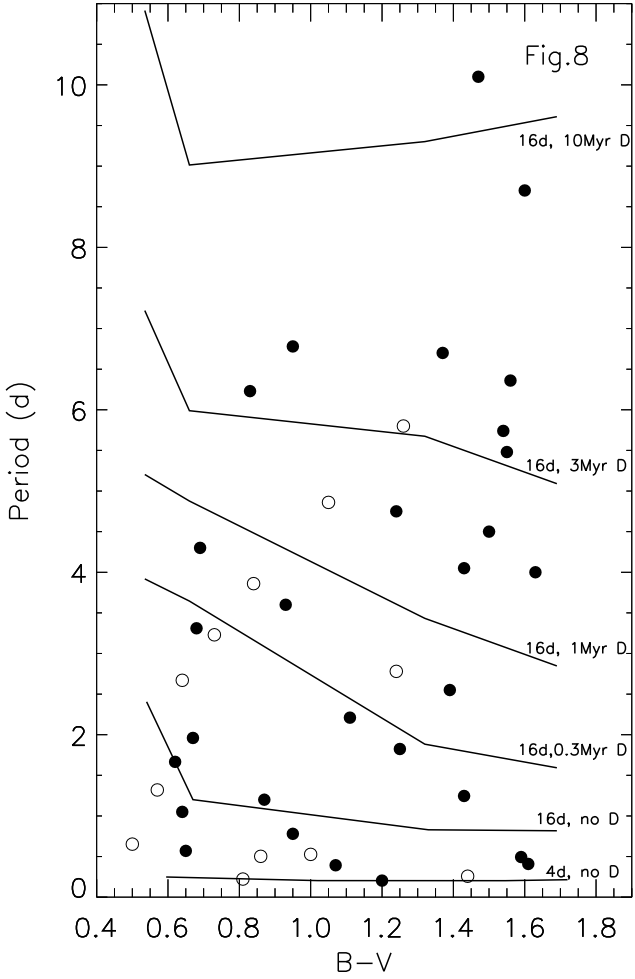}{8in}{0}{100}{100}{-310}{-50}
\caption[fig8.ps]{Rotation periods of stars in IC 2391 (open circles) and IC 2602 (filled circles) with theoretical models from Barnes (1998) overplotted. The initial periods and disk-lifetimes for the models are indicated on the right. The models suggest that disk-interaction is the norm rather than the exception.}\label{fig8}
\end{figure}

\end{document}